\documentclass[showpacs,floatfix,a4paper,prl,twocolumn]{revtex4}
\usepackage{amssymb}
\usepackage{amsfonts}
\usepackage{amsmath}
\usepackage{graphicx}
\usepackage{latexsym,epsfig,bm,times,psfrag,subfigure}
\usepackage{bm,times}
\usepackage{dcolumn}

\begin{document}

\title{Multicomponent Skyrmion lattices and their excitations}
\author{D.L.\ Kovrizhin$^{1,2}$, Beno\^{\i}t Dou\c{c}ot$^{3}$, and R.\
Moessner$^{1}$}
\affiliation{$^{1}$Max Planck Institute for the Physics of Complex Systems, N\"{o}%
thnitzer Str. 38, 01187 Dresden, Germany}
\affiliation{$^2$ Russian Research Centre, Kurchatov institute, 1 Kurchatov sq., 123098,
Moscow, Russia}
\affiliation{$^3$LPTHE, Universit\'e Pierre et Marie Curie-Paris 6 and CNRS UMR 7589,
Boite 126, 4 Place Jussieu, 75252 Paris Cedex 05}

\begin{abstract}
We study quantum Hall ferromagnets with a finite density topologically
charged spin textures in the presence of internal degrees of freedom such as
spin, valley, or layer indices, so that the system is parametrised by a $d$%
-component complex spinor field. In the absence of anisotropies, we find
formation of a hexagonal Skyrmion lattice which completely breaks the
underlying $SU(d)$ symmetry. The ground state charge density modulation,
which inevitably exists in these lattices, vanishes \emph{exponentially} in $%
d$. We compute analytically the complete low-lying excitation spectrum,
which separates into $d^{2}-1$ gapless acoustic magnetic modes and a
magnetophonon. We discuss the role of effective mass anisotropy for $SU(3)$%
-valley Skyrmions relevant for experiments with AlAs quantum wells. Here, we
find a transition, which breaks a six-fold rotational symmetry of a
triangular lattice, followed by a formation of a square lattice at large
values of anisotropy strength.
\end{abstract}

\maketitle

\emph{Introduction.}~Skyrmions in quantum Hall ferromagnets \cite{sondhi}
present an early example of topological excitations in condensed matter
physics, a topic of persistently large interest, which by now has grown to
encompass itinerant magnets with spin-orbit coupling \cite{nagaosa,bogdanov}
and spinor condensates \cite{cherng}. In the quantum Hall effect (QHE),
these topological textures carry a quantised charge, endowing them with
stability and providing possible probes via Coulomb interactions and charge
transport. The low-energy modes of these defects have possibly been observed
in NMR measurements \cite{barrett}. The textures are also exceptionally
tunable, as their density, and hence the relative strength of interactions,
can be modified by a gate potential or magnetic field. The possibility of
controlling spin by coupling to electron charge makes these systems
interesting candidates for spintronics applications.

In this work, we discuss the physics of Skyrmions with an enlarged internal
space. This space minimally consists of two components which are, most
simply, represented by physical up and down spins -- but other degrees of
freedom frequently play a role. Early examples are the layer index in double
quantum wells of GaAs, where interlayer phase coherence spontaneously
develops \cite{murphy,girvin}, or the valley degeneracy in semiconductors 
\cite{eng}. Recently, there has been a further proliferation, in the form of
spin, layer and valley degeneracies in graphene \cite{yoshioka}, and cold
atoms \cite{cherng}, where $SU(N)$ internal degrees of freedom and Skyrmions
have emerged as interesting topics.

At finite density of defects and in presence of Coulomb interactions,
Skyrmions can form a crystal (SC), analogous to Skyrme crystals of nuclear
physics \cite{Klebanov}. The properties of SCs have been extensively
discussed in the $SU(2)$ case, and a large number of results in a wide variety
of settings have been obtained using Hartree-Fock (HF) theory for $SU(4)$ systems
\cite{cote}. Most of these studies are based
on the numerical solution of the HF equations of motion, being hard to
perform due to presence of several lengthscales.

This motivates us to develop an \emph{analytical} approach to the general
case of Skyrmion crystals with an internal space parametrised locally by a $%
d $--component complex spinor. The simplifying feature of our model is that
the spinors live in the complex projective space $CP^{d-1}$, which supports
exact multi-Skyrmion solutions irrespective of the value of $d$, and, most
crucially, the assumption of large value of spin stiffness.

Our programme consists of two parts. First, we derive a set of largely
analytical results on the case of Skyrmion crystals which are described
locally by $SU(d)$ spinors. We find that macroscopic degeneracy, present in
the noninteracting case \cite{rajaraman}, is lifted in favour of a hexagonal
SC with the amplitude of charge density modulations decaying exponentially
with $d$. The excitation spectrum, obtained by studying small perturbation
of the crystal, exhausts the number of gapless modes consistent with the
original degeneracy. Here, $d^{2}-1$ of these modes correspond to generators
of the $SU(d)$ Lie algebra, in accordance with the Goldstone theorem,
supplemented by a magnetophonon excitation, appearing as a result of broken
translational invariance. The magnetophonon has a dispersion reminiscent of
a 2D Wigner crystal in a magnetic field \cite{fukuyama}. 
\begin{figure}[b]
\epsfig{file=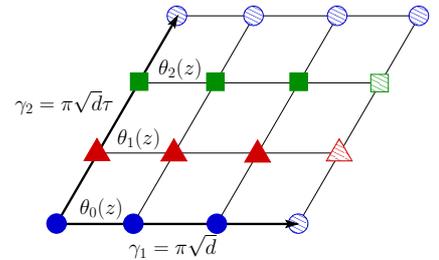,width=5.5cm}
\caption{(Color online). Positions of zeros of the basis theta-functions $%
\protect\theta _{p}(z)$ within a single unit cell of a Skyrmion lattice in
the $SU(3)$ case. The lattice vectors are indicated by arrows.}
\end{figure}

Second, as an application of our theory, we consider explicitly a model of
multi-component quantum Hall ferromagnet, motivated by experiments on
semiconductor quantum wells. The effects of anisotropies, present in these
systems, which break $SU(d)$ symmetry, are studied. With increasing $d$, the
number of possible types of anisotropies grows quickly along with the $%
d^{2}-1$ generators of $SU(d)$, and we concentrate on the simplest
nontrivial, yet instructive, and hitherto unstudied case of $SU(3)$
Skyrmions. These are of experimental interest for quantum wells grown in a
[111] direction, in particular a nascent effort on AlAs, where one finds the
appearance of three almost~degenerate valleys \cite{eng}. The latter are
subject to a `nematic' anisotropy recently discussed in the context of $%
SU(2) $ valley degrees of freedom \cite{parameswaran}, which we generalise
to the case of an arbitrary number of valleys. We show how this anisotropy
changes the nature of the crystalline ground state by combining spontaneous
symmetry breaking with the externally imposed one, leading to a collection
of different crystal structures. We also determine the gaps appearing in the
low-energy excitation spectrum.

The remainder of the paper is organised as follows. First, we introduce the $%
CP^{d-1}$ model describing spatial variations of internal degrees of
freedom, taking into account interactions between electrons. An exact HF
ground state in terms of spinors living in an analytic subspace is
presented. Starting from this state we derive linearized equations of motion
describing excitations of Skyrmion lattices. A generalization of Bloch
theorem leads, in the isotropic case, to a reduction of a system of $2d^{2}$
equations into $d^{2}$ decoupled pairs of linear equations, which analytical
solution provides the spectrum together with the eigenfunctions of the
excitations. Next, we consider a model of semiconductor, possessing three
degenerate valleys, which takes into account the `nematic' anisotropy, and
present results for the phase diagram and excitations.

\emph{The model.}~We consider a quantum Hall ferromagnet at filling factor $%
\nu \sim 1$ with internal degrees of freedom described by a $d$-component
spinor $w_{n}(\mathbf{r})$, which corresponds to the Slater determinant
state $|\Psi \rangle =\Pi _{\mathbf{r}}[\sum_{n=1}^{d}w_{n}(\mathbf{r})\hat{c%
}_{n}^{\dagger }(\mathbf{r})]|0\rangle ,$ where the spinor is normalized as $%
\sum_{n=1}^{d}|w_{n}(\mathbf{r})|^{2}=1$ at every point $\mathbf{r}$, $\hat{c%
}_{n}^{\dagger }(\mathbf{r})$ creates an electron in the $n$-th spin
component in the lowest Landau level (LLL) eigenstate with position $\mathbf{%
r}$. The latter are given in the coherent state representation, and $%
|0\rangle $ is the electron vacuum. For example, in the case of a bilayer
quantum Hall system, components of the spinor correspond to layer and spin
indices, and in graphene to spin and valley. The Hamiltonian of the system
is given by two contributions $H=H_{\mathrm{CP}}+H_{\mathrm{int}}$, where $%
H_{\mathrm{CP}}$ is proportional to the spin stiffness $\rho _{S}$, and $H_{%
\mathrm{int}}$ describes Coulomb interactions. A generic form of $H_{\mathrm{%
CP}}$, invariant under $SU(d)$ rotation and containing only first order
gradients reads \cite{moon,girvin}

\begin{equation}
H_{\mathrm{CP}}=2\rho _{S}\int d^{2}\mathbf{r\ }[\frac{(\nabla w,\nabla w)}{%
(w,w)}-\frac{(w,\nabla w)(\nabla w,w)}{(w,w)^{2}}],
\end{equation}%
where the integral is taken over the two dimensional plane, and the scalar
product is defined as $(a,b)\equiv \sum_{n}a_{n}^{\ast }(\mathbf{r})b_{n}(%
\mathbf{r})$. The Hamiltonian has a local \textquotedblleft
gauge\textquotedblright\ symmetry, in other words, multiplying a spinor by
an arbitrary function, $\tilde{w}_{a}(\mathbf{r})=h(\mathbf{r})w_{a}(\mathbf{%
r})$, does not affect the energy. The electron interactions with a potential 
$V(\mathbf{r}-\mathbf{r}^{\prime })$ are described by 
\begin{equation}
H_{\mathrm{int}}=\frac{1}{2}\int d^{2}\mathbf{r}\ d^{2}\mathbf{r}^{\prime }\
V(\mathbf{r}-\mathbf{r}^{\prime })Q(\mathbf{r})Q(\mathbf{r}^{\prime }),
\end{equation}%
in the lowest order of the gradient expansion. Here $Q(\mathbf{r})$ is the
topological charge density, which is proportional to the electronic charge
density in the quantum Hall effect \cite{sondhi,moon}, and is related to a
Pontryagin index of the normalized spinor field \cite{rajaraman}. In the
following we consider textures in the form of Skyrmion crystals and fix the
topological charge, given by the integral of $Q(\mathbf{r})$ inside the unit
cell (u.c.), to $2\pi d$. The two lattice vectors of the unit cell, which
define a parallelogram with opening angle $\phi ,$ can be chosen as $\gamma
_{1}=\pi \sqrt{d}$ and $\gamma _{2}=\pi \sqrt{d}\tau $, where $\tau
=e^{i\phi }$ (real/imaginary parts of $\gamma _{1,2}$ correspond to $x,y$
components of the vectors). We measure the lengths in units of $2l_{B}/\sqrt{%
|\nu -1|\mathrm{Im}\tau }$.

The energy $H_{\mathrm{CP}}$ is minimized by spinors, with entries analytic
functions of $z=x+iy$, whose number of zeros inside the u.c.~fixes the
topological charge sector. Without interactions there is a macroscopic
degeneracy coming from the arbitrariness in the choice of positions for
these zeros. In presence of $H_{\mathrm{int}}$ this degeneracy is lifted,
which can lead to formation of a Skyrmion crystal. It is impossible to write
an analytic expression for the ground state in the interacting case, thus we
propose a variational approach in which the trial state is still expressed
in terms of analytic functions $w_{a}(z)$. Although the Euler-Lagrange
equations, derived from $H_{\mathrm{CP}}$, are highly nonlinear, our
procedure is reminiscent of the LLL projection, familiar from the theory of
Abrikosov vortex lattices and the QHE physics. Our conjecture is that the
zero modes of the Hessian of $H_{\mathrm{CP}}$, spanning the analytic
subspace, are separated from the higher (non-analytic) ones by a spectral
gap of the order $\rho _{S}$, because the average topological charge
density, analogous to a magnetic field strength, is independent of $d$.
Further, the energy scale associated with the Coulomb interaction is $\rho
_{S}\sqrt{n_{S}}l_{B}$, where $n_{S}$ is the Skyrmion density and $l_{B}$
describes the magnetic length. This scale is smaller than $\rho _{s}$
provided that $n_{S}l_{B}^{2}\ll 1$, i.e.~if the filling factor remains
close to one. This justifies our variational treatment.

\emph{The method.}~We choose a basis of analytic functions defined by their
quasi-periodicity $\theta (z+\gamma )=e^{a_{\gamma }z+b_{\gamma }}\theta
(z),~\gamma =n_{1}\gamma _{1}+n_{2}\gamma _{2},$ where $a_{\gamma
},b_{\gamma }$ are complex numbers and $n_{1,2}\in Z$. Using the properties
of these functions under translations $\hat{T}_{\gamma _{2}/d}\theta
_{p}(z)=e^{-i\pi \tau (d+1)/d}\theta _{p+1}(z)\ $and $\hat{T}_{\gamma
_{1}/d}\theta _{p}(z)=e^{2i\pi p/d}\theta _{p}(z)$, we obtain the basis of
linearly-independent theta-functions with a given topological charge inside
the u.c.~%
\begin{equation}
\theta _{p}(z)=\sum\nolimits_{n=-\infty }^{\infty }e^{i\pi \tau d(n-\frac{p}{%
d})(n-1-\frac{p}{d})+2i\sqrt{d}(n-\frac{p}{d})z},
\end{equation}%
where $p$ runs from $0$ to $d-1$, in agreement with Riemann-Roch theorem 
\cite{debarre}. From the numerical minimization procedure, which was further
supported by a linear stability analysis, we find that the energy minimum is
achieved for a spinor formed from basis functions with equal amplitudes (up
to a global $SU(d)$ rotation) $w_{a}(z)=(\theta _{0},\ldots ,\theta _{d-1})$%
, and $\tau =e^{i\pi /3}$. This corresponds to a charge distribution with
the full symmetries of a hexagonal lattice, although the individual spinor
components possess a lower symmetry.

In order to calculate the excitations spectra of the SC we use the basis
formed from a set of quasi-periodic analytic functions $\chi _{p\mathbf{k}%
}(z)$ obtained from $\theta _{p}(z)$ by translations by vector $\mathbf{k}$
with components $2\pi m_{1}/N_{1}$ and $2\pi m_{2}/N_{2}$, where $%
m_{1},m_{2}\in Z$ and $N_{1}$ and $N_{2}$ is the number of unit cells on a
torus. An arbitrary analytic spinor can be written in terms of $\chi _{p%
\mathbf{k}}(z)$ as 
\begin{equation}
w_{a}(z)=\sum\nolimits_{\mathbf{k\in BZ}}\sum\nolimits_{b=1}^{d}\mathcal{U}%
_{ab}(\mathbf{k})\chi _{b\mathbf{k}}(z),\text{\ \ \ }a=0..d-1
\end{equation}%
see \cite{haldane} and a related work \cite{shlyapn}, where an analogous
approach was formulated to a $U(1)$ case for the problem of vortex lattices
in Bose-Einstein condensates. Here we introduce arbitrary complex square
matrices $\mathcal{U}$ of coefficients with linear dimension $d.$ We
consider time-dependent HF, with dynamics generated by a Lagrangian given as
a sum of the energetic term $H_{\mathrm{int}}$ and a Berry's phase term,
which is first order in time derivatives of the spinor components. Variation
of the Lagrangian with respect to parameters $\mathcal{U}_{ab}(\mathbf{k})$
gives a system of coupled equations describing the motion of a Skyrmion
lattice, constrained to the analytic subspace%
\begin{multline}
\int d^{2}\mathbf{r\ }\frac{\chi _{b\mathbf{k}}^{\ast }(z)}{(w,w)}\{{i\frac{%
\partial w_{a}}{\partial t}-i\frac{(w,\partial _{t}w)}{(w,w)}w_{a}(z)} \\
{-4w}_{a}(z){\int \Delta _{\mathbf{r}}V(\mathbf{r}-\mathbf{r}^{\prime })Q(}%
\mathbf{r}{^{\prime })d^{2}\mathbf{r}^{\prime }\}=0.}  \label{GP}
\end{multline}%
Equation (\ref{GP}) represents one of the main results of our paper.

\begin{figure}[bp]
\epsfig{file=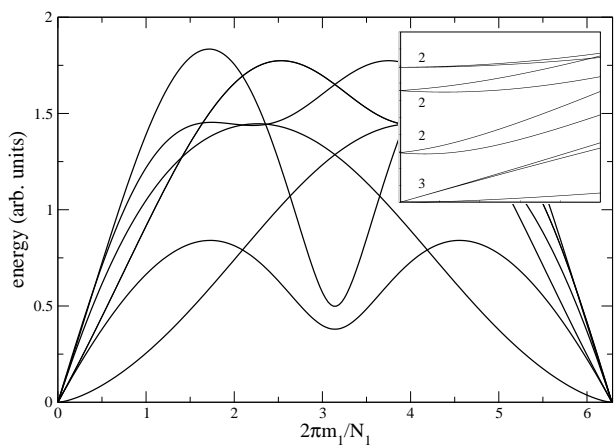,width=8cm}
\caption{Excitation spectra of a $SU(3)$ Skyrmion lattice, in the case of
Coulomb interactions, along a diagonal direction in the Brillouin zone, $%
k_{1}=k_{2}$. The lowest lying mode at small $k_{1}$ is a
magnetophonon. All remaining modes are linear gapless acoustic excitations.
(inset) appearance of gaps in the $SU(3)$ case at finite anisotropy strength.}
\end{figure}

\emph{LSWT and Bloch theorem.}~The system of equations (\ref{GP}) is very
general and is not amenable to analytic treatment, mostly because the
velocity field $\partial w_{a}/\partial t$ is not explicit. However, due to
the high degree of symmetry of our reference SC state, it is possible to
obtain full analytical results at arbitrary $d$ for small perturbations of
the lattice, i.e.~for linearized equations of motion. The matrices of
coefficients now can be written as a sum $\mathcal{U}_{ab}(\mathbf{k}%
)=\delta _{ab}\delta _{\mathbf{k},0}+u_{ab}(\mathbf{k}),$ where $u_{ab}(%
\mathbf{k})$ are small. The quasi-periodicity of theta-functions and the
periodicity (with elementary periods $\gamma _{1}/d$ and $\gamma _{2}/d$) of
the topological charge density of the SC leads to a generalized Bloch
theorem, and the system of equations fully decouples into $d^{2}$ pairs of
linear equations. These are readily solved, resulting e.g.~in the excitation
spectrum for a $SU(3)$ case presented in Fig. 2.

\emph{Results.}~The magnetophonon mode at small energies has a dispersion $|%
\mathbf{k}|^{\alpha }$, with $\alpha \sim 3/2$ in the case of Coulomb
interactions, like the dispersion of a 2D Wigner crystal in a magnetic
field. For a short-range interaction $\alpha \sim 2,$ while the spectrum is
linear at small energies, $\alpha \sim 1$, for logarithmic (2D Coulomb)
interactions. In the large-$d$ limit, topological charge density variations
of a SC have an exponentially small amplitude, for example for a square
lattice we find 
\begin{equation*}
Q(\mathbf{r})=2/\pi -4d\ e^{-\frac{\pi d}{2}}[\cos (2\sqrt{d}x)+\cos (2\sqrt{%
d}y)]+O(e^{-\pi d}),
\end{equation*}%
Coulomb interactions suppress charge density fluctuations, and, for a large
number of components, it becomes possible to arrange for a nearly-uniform
twist of the spinor. Interestingly, similar questions arise in the studies
of $SU(N)$ superconductors \cite{moore}.

\emph{SU(3) case.}~Let us discuss application of our general methods to a
potentially realizable case of Skyrmion crystals in AlAs quantum wells,
where the spin degree of freedom is quenched due to applied magnetic field,
and the valley-degeneracy leads to the three-component spinors with
coefficients $\alpha _{i}$, which we describe in zeroth order by a $SU(3)$
symmetric theory. Effective mass anisotropy within the valleys with relative
angle $\varphi $ generates a symmetry-breaking term in the Hamiltonian
describing a `nematic' anisotropy

\begin{equation}
H_{\mathrm{N}}=-\Delta _{0}c_{1}+2\Delta _{0}\varkappa \sum\nolimits_{i\neq
j}|\alpha _{i}|^{2}|\alpha _{j}|^{2},
\end{equation}%
where $\Delta _{0}=\sqrt{\pi /8}e^{2}/\varepsilon l_{B}$ is the exchange
energy in the isotropic case, and $l_{B}=\sqrt{\hbar c/eB}$ is the magnetic
length. With $K(x)$ denoting the complete elliptic integral of the first
kind, and $\varkappa =(c_{1}-c_{2})/2,$ the coefficients are given by 
\begin{equation}
c_{1}=\frac{2}{\pi }\frac{K(1-1/\lambda ^{2})}{\sqrt{\lambda }},\ \ \ c_{2}=%
\frac{2}{\pi }\frac{K(1-b^{2}/a^{2})}{a},
\end{equation}%
where $a^{2}=\lambda \cos ^{2}\tfrac{1}{2}\varphi +\lambda ^{-1}\sin ^{2}%
\tfrac{1}{2}\varphi $ and$\ b^{2}=\lambda \sin ^{2}\tfrac{1}{2}\varphi
+\lambda ^{-1}\cos ^{2}\tfrac{1}{2}\varphi $, 
\begin{figure}[t]
\epsfig{file=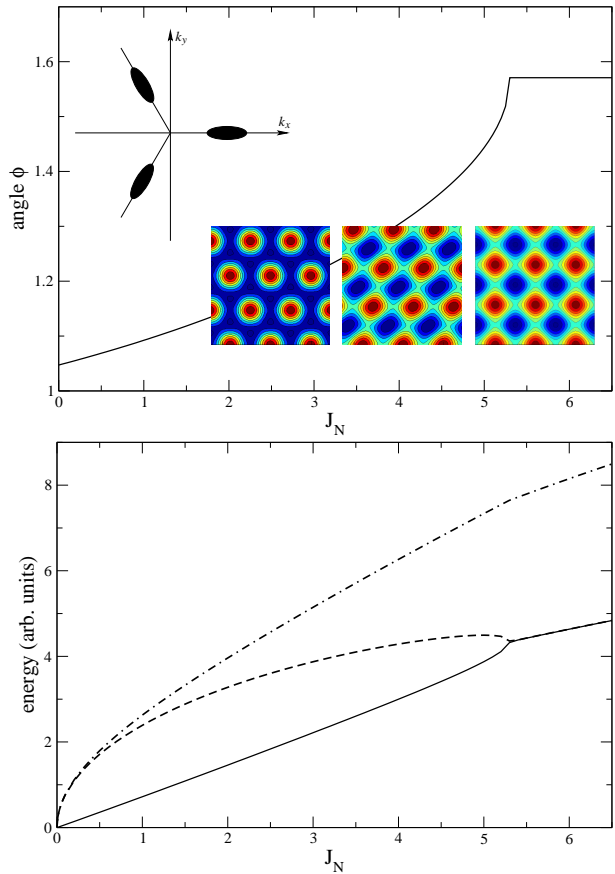,width=8cm}
\caption{a) Angle $\protect\phi $ in the $SU(3)$ case as a function of
nematic anisotropy strength. The crystal is hexagonal in the isotropic case
and becomes square at large $J_{N}=\varkappa \exp (\protect\pi d/2)/\protect%
\pi \protect\sqrt{2\protect\pi }|\protect\nu -1|^{3/2}$. b) Scaling of gaps
with nematic anisotropy strength. The first gap shows linear behavior, while
the other two increase as a square root. (upper inset of Fig.a) Schematic
picture of the valleys in AlAs quantum wells in the [111] direction showing
the three-fold valley degeneracy. The effective mass within each valley is
anisotropic with $m_{x}/m_{y}\approx 4.$ (lower inset of Fig.a) topological
charge densities of a Skyrmion crystal from weak to strong anisotropy.}
\label{figure4}
\end{figure}
and $\lambda =(m_{x}/m_{y})^{1/2}$ stands for the anisotropy parameter.
Notice that $\varkappa \geq 0$, describing repulsive interaction between the
different components $\alpha _{i}$ of the spinor. This reduces to the form
derived in \cite{parameswaran} for the $SU(2)$ case. In the isotropic limit, 
$m_{x}=m_{y}$, the ground state, for a given filling factor, is a triangular
Skyrmion lattice, which transforms into a square one with increasing value
of the anisotropy strength, see Fig. 3. This transition is due to the
relative scaling of Coulomb and nematic energies with $\mathrm{Im}\tau$. We
do not find the nematic phase transition of \cite{parameswaran} even at
large values of anisotropy strength. Note that although valley polarization is favoured
by the anisotropic term, its Coulomb energy cost turns out
to be so large that the different spinor amplitudes in our
variational ground-state remain equal, independently of the
value of $\varkappa$. 

Finite gaps appear in the spectrum in presence of anisotropy. The first
three modes remain gapless; they correspond to magnetophonon and to the two
remaining continuous symmetry generators associated with a two-dimensional
Cartan subgroup of $SU(3)$. The other six modes are now gapped, so that the
collective spectrum forms the pattern shown in Fig.3. The first two gaps
scale linearly with the anisotropy strengths, while the behavior is $\sim 
\sqrt{J_{N}}$ for the other four.

\emph{Energy scales.}~In the SU(3) valley-degenerate case we obtain at small
anisotropies $\mathcal{A}_{N}\equiv \Delta _{0}\varkappa /2\pi l_{B}^{2}\sim 
\frac{9}{128}\frac{\Delta _{0}}{2\pi l_{B}^{2}}(\lambda -1)^{2}$, and $\rho
_{S}=e^{2}/16\sqrt{2\pi }\varepsilon l_{B}\sim 5.2~\mathrm{K},$ with $\Delta
_{0}\sim 131~\mathrm{K}$. The anisotropy energy $\Delta _{0}\varkappa \sim
2.5~\mathrm{K}$ is much smaller than the spin stiffness scale $4\pi \rho
_{S}\sim 65~\mathrm{K,}$ where we assumed $\varepsilon =10\varepsilon _{0}$
for AlAs at $\nu =1$ and electron density $2.5\times 10^{11}\mathrm{cm}^{-1}$
\cite{shayegan}. The characteristic Coulomb energy is $\sim \rho _{S}|\nu
-1|^{1/2}$ and can be made small compared to the stiffness by changing the
filling factor. For relatively small values of $|\nu -1|\sim 0.1$ we
estimate the\ value of dimensionless parameter $J_{N}\sim 13.5$ which
corresponds to a square lattice, see Fig.3.

\emph{Conclusions}.~We have studied $SU(d)$ quantum Hall ferromagnets with
finite density of topological defects (Skyrmions). In presence of Coulomb
interactions the ground state of the system is a triangular Skyrmion
lattice, which breaks all the symmetries of the internal $SU(d),$ as well as
translational symmetry, generating $d^{2}$ Goldstone modes, $d^{2}-1$ of
which correspond to breaking of the $SU(d)$, and the remaining one being a
magnetophonon mode. We have explored the phase diagram of a $SU(3)$ QHE
ferromagnet, which is expected to be relevant to semiconductor
nanostructures with valley-degeneracies \cite{eng}. The lattice tilts
continuously as a function of anisotropy strength and becomes square through
a phase transition at a critical anisotropy value. In this case, three of
the excitation branches remain gapless, while others acquire gaps exhibiting
different scaling with anisotropy strength.

It would be interesting to extend our theory to the case of large, but
finite, spin-stiffness, and to derive an effective sigma-model for the
Skyrmion lattice in the projected subspace. The difficulty which one
immediately faces is that long-wavelength $SU(N)$ rotations are not
preserved under projection onto the analytic subspace. Another important
question is related to entanglement of internal degrees of freedom, as
studied for simple Skyrmions structures in \cite{Doucot}.

\emph{Acknowledgements}\textit{.~}We are grateful to J.~T.~Chalker for
insightful comments. D.~K.~acknowledges hospitality of the Theoretical
physics department at Oxford.


\begin{thebibliography}{99}
\bibitem{sondhi} S.~L.~Sondhi, A.~Karlhede, and S.~A.~Kivelson,
E.~H.~Rezayi, Phys. Rev. B \textbf{47}, 16419 (1993).%

\bibitem{nagaosa} X.~Z.~Yu, Y.~Onose, N.~Kanazawa, J.~H.~Park, J.~H.~Han,
Y.~Matsui, N.~Nagaosa and Y.~Tokura, Nature \textbf{465}, 901 (2010); S.~M%
\"{u}hlbauer, B.~Binz, F.~Jonietz, C.~Pfleiderer, A.~Rosch, A.~Neubauer,
R.~Georgii, P.~B\"{o}ni, Science \textbf{323}, 915 (2009).

\bibitem{bogdanov} A.~Bogdanov and A.~Hubert, Journ. of Magn. and Magn.
Mater. \textbf{138}, 255 (1994); A.~Bogdanov, and U.~R\"{o}\ss ler, Phys.
Rev. Lett. \textbf{87}, 037203 (2001). J.~H.~Han, J.~ Zang, Z.~Yang,
J.-H.~Park, and N.~Nagaosa, Phys. Rev. B \textbf{82}, 094429 (2010).

\bibitem{cherng} R.~W.~Cherng and E.~Demler, Phys. Rev. A \textbf{83},
053613 (2011); Phys. Rev. A \textbf{83}, 053614 (2011).

\bibitem{barrett} R.~Tycko, S.~E.~Barrett, G.~Dabbagh, L.~N.~Pfeiffer, and
K.~W.~West, Science \textbf{268}, 1460 (1995).

\bibitem{murphy} S.~Q.~Murphy, J.~P.~Eisenstein, G.~S.~Boebinger,
L.~N.~Pfeiffer, and K.~W.~West, Phys. Rev. Lett. \textbf{72}, 728 (1994).

\bibitem{girvin} S.~Girvin, \textquotedblleft The Quantum Hall Effect: Novel
Excitations and Broken Symmetries\textquotedblright , arXiv:cond-mat/9907002.

\bibitem{eng} S.~Prabhu-Gaunkar, S.~Birner, S.~Dasgupta, C.~Knaak, and
M.~Grayson, Phys. Rev. B \textbf{84}, 125319 (2011); K. Eng, R. N.
McFarland, B.~E.~Kane, Phys. Rev. Lett. 99, 016801 (2007); F.~Herzog,
M.~Bichler, G.~Koblm\"{u}ller, S.~Prabhu-Gaunkar, W.~Zhou, M.~Grayson,
Appl. Phys. Lett. \textbf{100}, 192106 (2012).

\bibitem{yoshioka} Y.~Sakurai and D.~Yoshioka, Phys. Rev. B \textbf{85},
045108 (2012); M.~O.~Goerbig, Rev. Mod. Phys. \textbf{83}, 1193 (2011).

\bibitem{Klebanov} I.~Klebanov, Nuclear Physics B \textbf{262}, 133, (1985);
T.~H.~R.~Skyrme, Proc. Roy. Soc. \textbf{260}, 127 (1961).

\bibitem{cote} R.~C\^{o}t\'{e}, D.~B.~Boisvert, J.~Bourassa, M.~
Boissonneault, and H.~A.~Fertig, Phys. Rev. B \textbf{76}, 125320 (2007).

\bibitem{rajaraman} R.~Rajaraman, Solitons and Instantons North-Holland,
Amsterdam, 1982; A.~M.~Polyakov, Gauge Fields and Strings, Harwood Academic
Publishers, 1987; Z.~F.~Ezawa, Quantum Hall Effects: Field Theoretical Approach and Related Topics,
World Scientific, 2008.

\bibitem{fukuyama} H.~Fukuyama, Solid State Comm., v. \textbf{17}, 10, 1323
(1975); R.~C\^{o}t\'{e} and A.~H.~MacDonald, Phys. Rev. B \textbf{44}, 8759
(1991); A.~G.~Green, I.~I.~Kogan and A.~M.~Tsvelik,
Phys. Rev. B \textbf{54}, 16838 (1996).

\bibitem{parameswaran} D.~A.~Abanin, S.~A.~Parameswaran, S.~A.~Kivelson,
S.~L.~Sondhi, Phys. Rev. B \textbf{82}, 035428 (2010).

\bibitem{moon} K.~Moon, H.~Mori, Kun~Yang, S.~M.~Girvin, A.~H.~MacDonald,
L.~Zheng, D.~Yoshioka, Shou-Cheng~Zhang, Phys. Rev. B \textbf{51}, 5138
(1995). 

\bibitem{debarre} O.~Debarre, \textquotedblleft Complex tori and Abelian
varieties\textquotedblright , American Mathematical Society, Texts and
Monographs, (1999).

\bibitem{haldane} F.~D.~M.~Haldane, E.~H.~Rezayi, Phys. Rev. B. \textbf{31},
2529 (1985); G.~Eilenberger, Phys. Rev. \textbf{164}, 628 (1967).

\bibitem{shlyapn} S.~I.~Matveenko and G.~V.~Shlyapnikov, Phys. Rev. A 
\textbf{83}, 033604 (2011).

\bibitem{moore} M.~A.~Moore, T.~J.~Newman, A.~J.~Bray, and S-K.~Chin, Phys.
Rev. B \textbf{58}, 936 (1998).

\bibitem{shayegan} Y.~P.~Shkolnikov, S.~Misra, N.~C.~Bishop, E.~P.~De
Poortere, and M.~Shayegan, Phys. Rev. Lett. \textbf{95}, 066809 (2005).

\bibitem{Doucot} B.~Dou\c{c}ot, M.~O.~Goerbig, P.~Lederer, and R.~Moessner,
Phys. Rev. B \textbf{78}, 195327 (2008).
\end{thebibliography}
\end{document}